\documentclass[prd, onecolumn, nofootinbib,notitlepage,floatfix]{revtex4-1}

\usepackage{amsmath}
\usepackage{graphicx}
\usepackage{dcolumn}
\usepackage{bm}
\usepackage{epsfig}
\usepackage{amssymb,latexsym,mathrsfs}
\usepackage{graphicx}
\usepackage{color}
\usepackage{hyperref}

\hypersetup{
    colorlinks=true,
    linkcolor=red,
    citecolor=blue,
} 

\newcommand{\be}{\begin{equation}}
\newcommand{\ee}{\end{equation}}
\newcommand{\bs}{\begin{split}} 
\newcommand{\bea}{\begin{eqnarray}}
\newcommand{\eea}{\end{eqnarray}}

\newcommand{\al}{\alpha} 
 
\newcommand{\dl}{\delta}

\newcommand{\mpl}{M^2_{\rm  pl}} 
\newcommand{\lamt}{\lambda^3}
\newcommand{\hds}{h}

\begin{document}

\title{Well Tempered Cosmology: Scales} 

\author{Eric V.~Linder$^{1,2}$}
\affiliation{
$^1$Berkeley Center for Cosmological Physics \& Berkeley Lab, 
University of California, Berkeley, CA 94720, USA\\  
$^2$Energetic Cosmos Laboratory, Nazarbayev University, 
Nur-Sultan, 010000, Kazakhstan }

\begin{abstract} 
Well tempered cosmology provides a well defined path for obtaining 
cosmology with a low energy cosmic acceleration despite a high (Planck) 
energy cosmological constant $\Lambda$, through a scalar field 
dynamically canceling $\Lambda$. We explore relations between the 
mass scales entering the various Horndeski gravity terms, and focus 
on the cases of only one or only two mass scales, obtaining general 
solutions for the form of the action. The resulting cosmology can be 
natural and viable, and as one of the only paths to dealing 
with the cosmological constant problem it has a rationale to be a 
benchmark cosmology. 
\end{abstract}

\date{\today} 


{
\let\clearpage\relax
\maketitle
}

\section{Introduction}

The vast majority of theories proposed to explain current cosmic 
acceleration ignore the cosmological constant problem: that vacuum 
energy with energy density at the Planck scale or some other high 
energy phase transition should dominate the universe \cite{martin,burgess,padilla}. A low energy 
cosmological constant, or scalar fields, or even modified gravity 
do not generally address this critical issue. 

However there are scalar-tensor modified gravity theories that 
can be devised to remove a high energy cosmological constant, 
through dynamical cancellation by the scalar field, and replace it 
by low energy cosmic acceleration. These include self tuning 
\cite{tune1,tune2,tune3,fab5,tunesch,fabfab, 2111.11448,2101.00965,2012.01838,2003.04303} and well tempered \cite{temper1,1812.05480,tempds,tempmk,tempexp,temptel} cosmologies. Well 
tempered cosmology can in addition provide a full cosmic history, 
from radiation domination to matter domination to cosmic 
acceleration in a natural manner. 

Well tempered cosmology uses particular relations between 
the terms of the Horndeski gravity Lagrangian in such a way that 
the scalar field and cosmic evolution equations become degenerate 
at a de Sitter attractor, ``on shell''. Several studies have been 
made of these relations in special cases, i.e.\ assumptions about 
the functional form of one or another Lagrangian term \cite{temper1,1812.05480,tempds} 
(even for a Minkowski background \cite{tempmk}). Perhaps the most 
general solution was obtained through a series expansion method, 
giving general relations between Lagrangian terms, as long as the 
functions were of polynomial form \cite{tempexp}. 

While polynomials (and in particular monomials) are an arbitrary 
choice, a natural constraint is shift symmetry, which brings some 
protection against quantum loop corrections. In this case the 
action terms only depend on functions of the quadratic kinetic 
variable $X\equiv\dot\phi^2/2$ of the scalar field $\phi$, and 
additive terms linear in $\phi$ (``tadpoles''). There then tend 
to be three mass scales in the theory (apart from the bare Planck 
mass, which is taken to be unity): one associated with the tadpole 
in the potential, one with the tadpole in the running Planck mass, 
and the scale of low energy cosmic acceleration, i.e.\ the asymptotic 
Hubble constant of the de Sitter state. 

Here we examine cases where relations exist between these mass scales, 
aligning scales in the well tempered cosmology to leave only one or 
two free mass parameters. In addition, we seek to go beyond polynomial 
functional forms to derive general relations between the action terms 
that provide this cosmology. 

In Section~\ref{sec:eqs} we introduce the Friedmann and scalar field 
evolution equations in the well tempered Horndeski theory, and derive 
the relations between the free action functions under various conditions. 
We consider late time and early time behaviors in 
Section~\ref{sec:lateearly}, concluding in 
Section~\ref{sec:concl}.

\section{Evolution Equations and Action Relations} \label{sec:eqs} 

Horndeski gravity is the most general scalar-tensor theory 
delivering second order equations of motion, and is composed of 
three action terms (we take the  $G_5=0$ so that the 
speed of gravitational waves equals the speed of light), 
generalizing general relativity. Imposing shift symmetry, the 
three terms are 
\bea 
K(X,\phi)&=&K(X)-\lambda^3\phi\,,\\ 
G_3(X,\phi)&=&G_3(X)\,,\\ 
G_4&=&\frac{1}{2}\left(\mpl+M\phi\right)\,. 
\eea 
We see there are two functions of $X$ (we write $K(X)$ to denote 
the $X$ dependent part of the generalized kinetic term; since we 
never again use the notation $K(X,\phi)$ this should not be confusing) 
and two mass scales -- $M$ and $\lambda$ in the tadpoles for the 
Planck mass coupling and scalar linear potential respectively. 

For a flat Friedmann-Lema{\^\i}tre-Robertson-Walker cosmology, the 
evolution of the cosmic expansion rate $H(t)=\dot a/a$ and scalar 
field are given by coupled equations of motion: 
\bea 
3H^2(\mpl+M\phi)&=&\rho_m+\Lambda+2XK_X-K+6H\dot\phi XG_{3X} 
-3MH\dot\phi  \label{eq:fried}\\ 
-2\dot H\,(\mpl+M\phi)&=&
\rho_m+P_m+\ddot\phi\,(M-2XG_{3X})-H\dot\phi\,(M-6XG_{3X})
+2XK_X \label{eq:fulldh}\\ 
0&=&\ddot\phi\,\left[K_X+2XK_{XX}+6H\dot\phi(G_{3X}+XG_{3XX})\right]\notag\\ 
&\qquad&+3H\dot\phi\,K_X+\lambda^3+6XG_{3X}(\dot H+3H^2) 
-3M(\dot H+2H^2)\,,  \label{eq:fullddphi} 
\eea 
where a subscript $X$ denotes a derivative with respect to $X$, and 
$\rho_m$ and $P_m$ are the matter 
energy density 
and pressure respectively. 

Well tempering arises from degeneracy of the evolution equations 
``on shell'', in this case at a de Sitter asymptote 
$H=H_{\rm dS}\equiv h=\,$const. The de Sitter mass scale $h$ 
is the third mass scale in the system. 
The degeneracy is satisfied by 
equating the coefficient of $\ddot\phi$ in one of 
Eq.~\eqref{eq:fulldh} or Eq.~\eqref{eq:fullddphi} times 
the remaining terms in the other equation with the same product  
for the equation selection interchanged. 
That is, if $A\ddot\phi+B=0=C\ddot\phi+D$ holds at the de Sitter asymptote then we solve both equations if $AD=BC$, as seen by multiplying 
the equations by $C$ and $A$ respectively: $AC\ddot\phi+BC=AC\ddot\phi+AD$ when $AD=BC$. 
Thus the degeneracy 
condition is 
\bea 
& &(M-2g)\left\{3\hds\dot\phi K_X+18\hds^2 g-6\hds^2 M+\lamt\right\}\notag\\ 
& &\qquad= (K_X+2XK_{XX}+6\hds\dot\phi g_X) \left[-\hds\dot\phi(M-6g)+2XK_X\right]\,, \label{eq:degen} 
\eea 
where we simplify notation by writing $g\equiv XG_{3X}$. 

Various solutions to this equation for well tempering have been 
found in \cite{tempds,tempexp} for polynomial forms of $g$ or $K_X$. 
The nonlinear nature of the equation has not allowed a general 
solution. However here we will adopt  relations between mass 
scales that can allow for general solutions.

\subsection{Two Mass Scales ($\lambda^3=3h^2M$)} \label{sec:lam3m} 

An intriguing aspect of the degeneracy equation~\eqref{eq:degen}  
is that when $\lambda^3=3h^2M$ then the two terms $\{\,\}$ and 
$[\ ]$ are proportional. That is, when we relate the three mass 
scales in this particular way (one could also view this as 
$h^2=\lambda^3/(3M)$), leaving only two mass scales in the 
system, then the equation greatly simplifies to 
\be 
(M-2g)Z=\frac{\dot\phi}{3h}\,\left(K_X+2XK_{XX}+6\hds\dot\phi g_X\right)Z\,, 
\ee 
where $Z\equiv 3\hds\dot\phi K_X+18\hds^2 g-3\hds^2 M$, i.e.\ the 
$\{\,\}$ term (when $\lambda^3=3h^2M$). 
This then gives two cases of solutions, when $Z=0$ and $Z\ne0$. 

The condition 
$Z=0$ leads to  
\bea 
g&=&\frac{M}{6}-\frac{\sqrt{2}\,X^{1/2}K_X}{6h} \label{eq:gz0}\\ 
K_X&=&\frac{hM}{\sqrt{2}}X^{-1/2}-\frac{6h}{\sqrt{2}}X^{-1/2}g\ ; \label{eq:kz0}
\eea 
these are equivalent expressions, i.e.\ one can either specify $K(X)$ and 
derive $g$ or the other way around. Interestingly, on shell the $\dot H$ Friedmann 
equation~\eqref{eq:fulldh} becomes $0=\ddot\phi(M-2g)$, implying that at the de Sitter 
asymptote the field coasts, $\ddot\phi=0$. Note this is {\it not\/} 
the zero coefficient case called $(\checkmark)$ in \cite{tempds}; the 
equations of motion remain intact, and do not become trivial. 
(If we choose $g=M/2$, i.e.\ $K_X=-hM\sqrt{2}X^{-1/2}$, then they do 
become trivial.) The choice $\lambda^3=3h^2M$ is unique in the sense that with this relation, all $\phi$ dependence drops out of the Friedmann equation~\eqref{eq:fried} on-shell. We are left on-shell with an algebraic relation between $X$ and the vacuum energy $\Lambda$, which requires $X = {\rm constant}$, hence $\ddot{\phi} = 0$. 
This is a nice property, in that it has no runaway behavior. 

For $Z\ne0$, we can solve the degeneracy equation to give 
\bea 
g&=&\frac{M}{2}+cX^{-1/2}+\frac{X^{-1/2}}{6h\sqrt{2}}\,\left[K(X)-2XK_X\right] \label{eq:gz}\\ 
K_X&=&bX^{-1/2}+\frac{3hM}{2\sqrt{2}}\,X^{-1/2}\ln X-3h\sqrt{2}X^{-1/2}g-\frac{3h}{\sqrt{2}}X^{-1/2}\int\frac{dX}{X}\,g\ . \label{eq:kz}
\eea 
Again, these two expressions are equivalent. Here $b$ and $c$ are 
arbitrary constants, and recall that $K(X)$ refers to the $X$ dependent 
part of $K$, without the tadpole (since $K(X)$ arises from integrating $K_X$). 
The solution Eq.~\eqref{eq:kz} looks like the Branch B solution of 
\cite{tempexp}, but here it is valid for all $g(X)$, not simply a finite 
polynomial as was the case for that paper. 

We emphasize that we can apply the above well tempering conditions to 
general functional forms, such as algebraic expressions. 

From $g$ and $K_X$ we can readily obtain the action functions $G_3$ 
and $K$. For $Z=0$ these are 
\bea 
G_3&=&\frac{M}{6}\,\ln X-\frac{1}{3h\sqrt{2}}X^{-1/2}K(X)
-\frac{1}{6h\sqrt{2}}\int dX\,X^{-3/2}K(X) \label{eq:g30}\\ 
K&=&hM\sqrt{2} X^{1/2}-3h\sqrt{2}\int dX\,X^{-1/2}g-3h^2M\phi\,. \label{eq:k0}
\eea 
(Note that a constant in $G_3$ gives an ignorable total derivative in the 
action.) For $Z\ne0$ they are 
\bea 
G_3&=&\frac{M}{2}\,\ln X-2cX^{-1/2}-\frac{1}{3h\sqrt{2}}\,X^{-1/2}K(X) \label{eq:g3}\\ 
K&=&a X^{1/2}+\frac{3Mh\sqrt{2}}{2}\,X^{1/2}\ln X
-3h\sqrt{2}\int dX\,X^{-1/2}\left[g+\frac{1}{2}\int\frac{dX'}{X'}\,g\right]\,-3h^2M\phi\,, \label{eq:k}
\eea 
where $a$ is an arbitrary constant. 

From the Friedmann expansion equation~\eqref{eq:fried}, on substituting 
back in the expressions for $g$ we obtain 
\bea 
3H^2\mpl&=&\rho_m+\Lambda+2XK_X\left(1-\frac{H}{h}\right)-K(X)-2MH\sqrt{2}X^{1/2}+3M\phi(h^2-H^2)\qquad[Z=0]\\ 
3H^2\mpl&=&\rho_m+\Lambda+6cH\sqrt{2}-\left[K(X)-2XK_X\right]\left(1-\frac{H}{h}\right)+3M\phi(h^2-H^2)\,, \quad\qquad\ [Z\ne0]
\eea 
for the $Z=0$ and $Z\ne0$ cases respectively. Recall $K(X)$ has the 
tadpole $3h^2M\phi$ removed (and moved to the last term). Thus the 
field $\phi$ and its evolution $X$ can dynamically cancel a bare 
cosmological constant everywhere (in the $Z\ne0$ case, at the  
asymptotic limit $H=h$ most terms vanish but the free $c$ term can still 
cancel $\Lambda$). 

We can also check the soundness of the theory through the no ghost 
and Laplace stability criteria. For the no ghost condition, it is 
convenient to work with the property functions \cite{bellsaw}, for which 
\bea 
\al_M&=&\frac{M\dot\phi}{H(\mpl+M\phi)}\\ 
\al_B&=&\frac{\dot\phi(2g-M)}{H(\mpl+M\phi)}\\ 
\al_K&=&\frac{2X(K_X+2XK_{XX})+12H\dot\phi Xg_X}{H^2(\mpl+M\phi)}\ , 
\eea 
for the Planck mass running, braiding, and kineticity respectively. 
When $Z=0$, 
\be 
\al_K=\frac{2X(K_X+2XK_{XX})(1-H/h)}{H^2(\mpl+M\phi)}\ ,  
\ee 
i.e.\ the kineticity vanishes at the de Sitter asymptote (the 
same happens in quintessence and k-essence). When $Z\ne0$, 
\bea  
\al_K&=&\frac{-3h\sqrt{2}X^{1/2}(2g-M)-12h\sqrt{2}X^{3/2}g_X(1-H/h)}{H^2(\mpl+M\phi)}\\ 
&=&\frac{2X(K_X+2XK_{XX})(1-H/h)-6Hc\sqrt{2}-(K-2XK_X)H/h}{H^2(\mpl+M\phi)}\ .
\eea  
In the de Sitter limit, $\al_K\to -3\al_B$. 

The no ghost condition is 
\be 
\al_K+\frac{3}{2}\al_B^2\ge0\,. 
\ee 
There is little we can say in general about this (or the Laplace 
stability condition, e.g.\ see Appendix~A.2 of \cite{tempds}), 
without adopting a specific $K$ or $G_3$. 
However, in the de Sitter limit, we can see that 
for the $Z=0$ case the theory is ghost free. 
In the $Z\ne0$ 
case, in the de Sitter limit the no ghost condition 
becomes $-(3\al_{B,{\rm dS}}/2)(2-\al_{B,{\rm dS}})\ge0$, so we 
require $\al_{B,{\rm dS}}\le0$ or $\al_{B,{\rm dS}}\ge2$.

\subsection{Two Mass Scales ($M=0$)} \label{sec:twom0} 

Let us consider instead two mass scales $\lambda$ and $h$, 
keeps the $G_4$ term 
at the minimal coupling of general relativity, and simplifies 
issues from nonminimal coupling such as discussed in 
Sec.~4 and Appendix~C of \cite{tempds}. The degeneracy equation 
becomes 
\be 
4Xg_X+2g\left(1+\frac{\lambda^3\dot\phi}{3h[2XK_X+6h\dot\phi g]}\right) 
+\frac{\dot\phi}{3h}\,(K_X+2XK_{XX})=0\,. \label{eq:degenm0} 
\ee 
We must have $Y\equiv[2XK_X+6h\dot\phi g]\ne0$ or the 
original degeneracy equation forces $\lambda=0$. Being nonlinear, 
this equation cannot be solved in general, but one solution is, for $Y=c$ with $c$ a constant, 
\bea 
K_X&=&\frac{\lambda^3}{3h\sqrt{2X}}\,\left(1+\frac{\lambda^3\sqrt{2X}}{3hc}\right)^{-1} \qquad , \qquad K=c\,\ln\left(1+\frac{\lambda^3\sqrt{2X}}{3hc}\right)-\lambda^3\phi \\ 
g&=&\frac{c}{6h\sqrt{2X}}-\frac{\lambda^3}{18h^2}\,\left(1+\frac{\lambda^3\sqrt{2X}}{3hc}\right)^{-1}\ . 
\eea 
Note these are algebraic functions, not covered by the 
polynomial solutions of \cite{tempds}. 

This provides a well tempered solution with scalar field equation 
on shell 
\bea  
\ddot\phi&=&3h\dot\phi+\frac{\lambda^3}{c}\,\dot\phi^2\\ 
\dot\phi&=&\frac{-3hc}{\lambda^3}\,\frac{Ae^{3ht}}{Ae^{3ht}-1}\,,  
\eea  
where $c<0$. Thus again $\dot\phi$ approaches 
a constant, freezing $X$. 
Note that while $c$ is effectively another mass scale (it has the 
same dimensions as $X\sim m^2/t^2$), we could define it 
as some combination of $\lambda$ and $h$, 
e.g.\ $c\sim -(\lambda^3/h)^2$.

\subsection{Single Mass Scale} \label{sec:m0} 

We can further reduce the number of free mass scales by setting 
$M=\lambda=0$. Since again  
the Planck mass does not run, this avoids 
the same problematic issues as mentioned in 
the previous subsection. 
This preserves all the results of Section~\ref{sec:lam3m}, while simplifying some, e.g.\ the Laplace stability condition 
for the $Z=0$ case in the de Sitter limit becomes 
\be 
|g|_{\rm dS}\le \frac{h\mpl}{\dot\phi} \qquad {\rm equiv.}\qquad |XK_X|_{\rm dS}\le 3h^2\mpl\ . 
\ee
The single mass scale is $h$,  which will be determined by 
the coefficients in $K$ or $G_3$, e.g.\ $g_n$ in  
$g=g_n X^n$.

\section{Asymptotic Behaviors} \label{sec:lateearly}

\subsection{Late Time de Sitter Attractor} 

To check the attractor behavior to the de Sitter asymptote, we 
perturb $H=h+\dl h$, $\phi=\phi_0+\dl\phi$. On shell, for the $Z=0$ case we have 
$\ddot\phi_0=0$. In the $M=\lambda=0$ single 
scale model, for simplicity, the solution to first order in the perturbations 
is 
\bea 
\dl h&=&\dl h_0\,e^{-3K_X ht}\\ 
\dl\ddot\phi&=&3\dot\phi_0\left(1+\frac{6\mpl h^2}{\dot\phi_0^2}\right)\,\dl h\sim e^{-3K_X ht}\ . 
\eea 
As long as $K_X\ge0$ in the de Sitter limit, the attractor 
is stable. Note that $\ddot\phi$ approaches zero from below, 
i.e.\ the kinetic term $X$ slows to a constant, with the field 
evolving as 
\be 
\dot\phi(t)=\dot\phi_0\,\left[1-\frac{\dl h_0}{hK_X}\left(1+\frac{6\mpl h^2}{\dot\phi_0^2}\right)\,e^{-3K_X ht}\,\right]\ . 
\ee

\subsection{Early Time Matter Dominated Behavior} 

We want to insure a viable cosmology at early times, having 
a standard matter dominated epoch before late time acceleration. 
Consider Eq.~\eqref{eq:fullddphi} at early times when $H\gg h$. 
Keeping the leading order terms in $H$ gives 
\be 
0\approx 6Hg_X\dot\phi\ddot\phi+9(1-w_b)H^2g-\frac{3}{2}(1-3w_b)H^2 M\, 
\ee 
where we have used $\dot H=-(3/2)(1+w_b)H^2$ for $w_b$ the 
early time background equation of state. This has the solution 
\be 
g\sim c\ \frac{1-w_b}{1+w_b}\,t^{-(1-w_b)/(1+w_b)}+\frac{M}{6}\frac{1-3w_b}{1+w_b}\,, 
\ee 
where $c$ is a constant. 
We see that $g$ gets large at early times. 
Through Eqs.~\eqref{eq:kz0} or \eqref{eq:kz} we see that this 
implies that $X^{1/2}K_X$ also is large at early times. 
From the Friedmann equation~\eqref{eq:fried} we see that 
$\dot\phi$ grows as $t^{2/(1+w_b)}$ so the field starts off slowly 
rolling. 
Thus, for 
a viable matter era we require $K_X\sim X^{<-1/2}$. 
This also guarantees freedom from ghosts at these early times, 
and that the property functions all go 
to zero (restoring general relativity) at early times.

\section{Conclusions} \label{sec:concl} 

As one of the few physical approaches that does not sweep the problem of a high energy cosmological constant under the rug, 
well tempered cosmology is an attractive avenue to deal with our universe, accelerating at a low energy scale. 
In well tempering, terms in the Horndeski action are related to each other, 
enabling a dynamical cancellation of the cosmological constant, 
while preserving matter and giving rise to low energy acceleration. 
Here we go further, and relate not just the forms of the 
action terms (while leaving them more general than previous solutions) but also their mass scales. 

We present three such relations: removing one mass scale 
by a unique relation $\lambda^3=3h^2M$, 
by setting $M=0$, or removing two mass scales by 
setting $M=0=\lambda$. Each has its own particular 
characteristics, benefits, and solutions. In the first case, one 
branch of the solution  
bounds the quadratic kinetic 
quantity $X=\dot\phi^2/2$, such that 
it goes to a constant in the de Sitter asymptotic 
rather than running away. This is related to 
$\phi$ (but not $X$) dependence dropping out of 
the Friedmann equation in that limit. 
The field behavior in the second branch depends on 
the particular form of $K(X)$ or $G_3(X)$. The second case can also 
give bounded solutions, and moreover 
possesses the minimal coupling of general relativity 
to the Ricci scalar. The third case has a blend of the above properties. 

While with three mass scales, Horndeski terms in 
the Friedmann equation partially cancel due to the 
degeneracy condition, giving a constant piece to 
dynamically cancel a large cosmological constant, 
while $\dot\phi$ runs away to infinity at late times. With only two 
mass scales, the constant piece can be obtained by 
freezing $X$ in the de Sitter limit, so $\dot\phi$ goes to a constant. 

In addition, we investigated the form of the property 
functions, more closely related to phenomenology, 
and the ghost free condition. Finally, we explored the late time de Sitter 
attractor nature, and the early time matter dominated 
behavior. 

The new general functional solutions, not limited 
merely to the previous polynomial forms, and 
the discovery of cases with bounded scalar field evolution, show that well tempered cosmology is a rich field. 
Dynamical cancellation of a high energy cosmological 
constant and protection against quantum radiative 
corrections (due to shift symmetry) make such an 
approach a leading ``benchmark'' to use in testing our low energy accelerating universe.

\acknowledgments  

I gratefully acknowledge Stephen Appleby for useful discussions. 
This work is supported in part by the Energetic Cosmos Laboratory and by the 
U.S.\ Department of Energy, Office of Science, Office of High Energy 
Physics, under contract no. DE-AC02-05CH11231.

\end{document}